\documentclass[aps,prd,twocolumn,nofootinbib,superscriptaddress,floatfix]{revtex4-2}
\usepackage[T1]{fontenc}
\usepackage[utf8]{inputenc}
\usepackage{amsmath,amssymb,bm,mathtools,graphicx,booktabs,multirow,array}
\usepackage[dvipsnames]{xcolor}
\usepackage[colorlinks=true,linkcolor=BrickRed,citecolor=BrickRed,urlcolor=BrickRed]{hyperref}
\usepackage[capitalize]{cleveref}
\usepackage{microtype}
\usepackage{orcidlink}
\usepackage{float}
\allowdisplaybreaks

\newcommand{\Tr}{\mathrm{Tr}}
\newcommand{\dd}{\mathrm{d}}

\newcommand{\be}{\begin{equation}}
\newcommand{\ee}{\end{equation}}
\newcommand{\bea}{\begin{align}}
\newcommand{\eea}{\end{align}}
\newcommand{\cO}{\mathcal{O}}

\begin{document}

\title{Post-Inflationary Quenched Production of Axion $SU(2)$ Dark Matter }

\author{Imtiaz Khan}
\email{ikhanphys1993@gmail.com}
\affiliation{Department of Physics, Zhejiang Normal University, Jinhua, Zhejiang 321004, China}
\affiliation{Research Center of Astrophysics and Cosmology, Khazar University, Baku, AZ1096, 41 Mehseti Street, Azerbaijan}
\affiliation{Zhejiang Institute of Photoelectronics, Jinhua, Zhejiang 321004, China}
\author{Pirzada}
\email{pirzada@itp.ac.cn}
\affiliation{CAS Key Laboratory of Theoretical Physics, Institute of Theoretical Physics, Chinese Academy of Sciences, Beijing 100190, China}
\affiliation{School of Physical Sciences, University of Chinese Academy of Sciences, Beijing 100049, China}
\author{G. Mustafa}
\email{gmustafa3828@gmail.com}
\affiliation{Department of Physics, Zhejiang Normal University, Jinhua, Zhejiang 321004, China}


\begin{abstract}

The relic abundance of vector dark matter originating from an inherited axion–SU(2) condensate is typically determined by implementing an adiabatic matching procedure across the symmetry-breaking transition. We demonstrate that this outcome does not arise in the generic case. The post-inflationary crossover can be formulated as a dynamical quantum quench problem, in which the residual coherent component of the field is characterized by a survival factor that induces an order-unity renormalization of the standard abundance relation. Expressed in conformal time, the spatially homogeneous condensate dynamics reduce to those of a canonical oscillator with quartic and quadratic self-interactions. This representation enables an analytic determination of the matching conditions across the symmetry-breaking transition, the derivation of the corresponding quench work and excess energy relations, and a quantitative validation of the coherent sector description via numerical simulations in both Minkowski and Friedmann–Robertson–Walker backgrounds. We also formulate the homogeneous fluctuation theory via the diagonal-\(SO(3)\) \(1\oplus3\oplus5\) decomposition and isolate a soft traceless-symmetric quintet with a \(k=0\) vacuum obstruction a regulated ultraviolet adiabatic bound, and a positive quartic stabilization term. Collectively, these results refine the theoretical description of inherited non-Abelian dark matter production and establish the necessary infrared framework for subsequent investigations of finite-\(k\) gauge–Higgs transfer dynamics.
\end{abstract}

\maketitle

\section{Introduction}\label{sec:intro}

The inflationary generation of vector dark matter has been investigated via several distinct classes of mechanisms, including production from inflationary quantum fluctuations, purely gravitational processes, dynamics associated with spontaneous symmetry breaking, and scenarios involving time-dependent effective masses \cite{graham_vector_2016,ema_production_2019,ahmed_gravitational_2020,sato_gravitational_2022,dror_parametric_2019,nakayama_yin_symmetry_2021,salehian_vector_2021,Pirzada_2026uak,kaneta_massvarying_2023}. A particularly well-motivated candidate is coherently oscillating hidden-photon dark matter produced via a misalignment-type mechanism. In its simplest realization, however, this scenario is impeded because the canonically normalized vector field typically acquires a Hubble-scale effective mass during inflation, while the most straightforward modifications tend to induce either phenomenological tensions or dynamical instabilities \cite{nelson_darklight_2011,arias_wispy_2012,nakayama_vcodm_2019,nakayama_constraint_2020}.The vector-dark-matter problem is therefore highly sensitive to the dynamical origin of the background field.

A qualitatively different resolution arises when inflation dynamically generates an isotropic non-Abelian gauge condensate that later undergoes symmetry breaking after inflation. In axion--SU(2) systems this isotropic background is sustained by the locking of spatial and gauge indices in the homogeneous configuration, a structure familiar from chromo-natural inflation and its extensions \cite{adshead_wyman_chromonatural_2012,dimastrogiovanni_peloso_stability_2013,adshead_martinec_wyman_perturbations_2013,adshead_sfakianakis_wyman_higgsed_2016,Barbon:2025wjl,Pirzada_2026uak,wolfson_maleknejad_komatsu_attractive_2020,wolfson_etal_isotropic_2021,ishiwata_komatsu_obata_backreaction_2022}. The resulting condensate evades the basic pathologies of a single homogeneous U(1) vector most notably the tension with statistical isotropy and the ghost problem of the original non-minimal constructions while retaining the coherent character that makes misalignment-type dark matter phenomenologically compelling \cite{nelson_darklight_2011,arias_wispy_2012,Pirzada:2026npl,nakayama_constraint_2020}. Once SU(2) is spontaneously broken after inflation, the inherited condensate becomes a coherently oscillating massive vector field.

This post-inflationary scenario sits at the intersection of several active lines of work. Axion--SU(2) inflation continues to provide a versatile setting for chiral gravitational waves, backreaction, primordial-black-hole production, and thermalization diagnostics \cite{adshead_martinec_wyman_perturbations_2013,fujita_imagawa_murai_2022,ishiwata_komatsu_obata_backreaction_2022,Khan:2025ibo,dimastrogiovanni_fasiello_papageorgiou_2024,Ijaz:2024zma,Ijaz:2023cvc,Muhammad:2026gmg,bhattacharya_fasiello_papageorgiou_dimastrogiovanni_2025}. In parallel, vector dark matter from time-dependent masses or symmetry breaking has been developed in conceptually different realizations \cite{dror_parametric_2019,nakayama_yin_symmetry_2021,salehian_vector_2021,kaneta_massvarying_2023}. Phenomenological viability is further sharpened by the mature dark-photon literature and by late-formation constraints on scenarios in which the dark sector becomes cold only after a delayed transition \cite{arias_wispy_2012,sarkar_das_sethi_2015,corasaniti_etal_2017,das_nadler_2021}. The issue is therefore not whether the inherited-condensate mechanism is interesting, but how robust its coherent abundance remains once symmetry breaking is treated as a genuinely dynamical stage.

A central simplifying assumption in the inherited-condensate scenario is adiabatic matching through symmetry breaking. In the post-inflationary scenario the gauge condensate initially behaves as a quartic SU(2) oscillator and later crosses over to a quadratic massive vector. The final abundance is inferred by assuming that the symmetry-breaking timescale is longer than the oscillation timescale, so that the relevant adiabatic invariant is conserved. This is a physically motivated approximation, and in the slow-transition regime it is expected to be correct. But the approximation also obscures the most delicate part of the mechanism: the post-inflationary transition from an inherited non-Abelian condensate to a massive-vector coherent state.

The present problem is also distinct from more familiar nonequilibrium settings. It is not a Kibble--Zurek transition, because the observable of interest is the coherent remnant of an already established condensate rather than the freeze-out density of defects \cite{kibble_topology_1976,zurek_cosmological_1985}. Nor is it a standard preheating problem, where nonadiabaticity is driven by a periodically pumped instability band \cite{kofman_linde_starobinsky_1994,greene_kofman_linde_starobinsky_1997}. Here the nonlinearity is intrinsic to the non-Abelian gauge sector, the transition is tied to symmetry breaking of the inherited condensate itself, and the primary quantity to control is the surviving coherent branch. Closely related nonadiabatic vector-dark-matter scenarios with symmetry breaking or time-dependent masses have been explored in other settings \cite{nakayama_yin_symmetry_2021,salehian_vector_2021,kaneta_massvarying_2023}; the focus here is the inherited non-Abelian condensate and the dynamical robustness of its post-transition coherent abundance.

{The coherent-sector analysis presented below is applicable to the kinetically resolved regime of the inherited gauge condensate. In this regime, the symmetry-breaking order parameter can be sourced or modulated by either a thermal or a nonthermal environment, while scattering processes that randomize the homogeneous \(SU(2)\) gauge condensate remain inefficient throughout the transition and thus proceed on parametrically longer time scales. Quantitatively, the relevant gauge-sector decoherence or thermalization rate $\Gamma_A$ satisfies $\Gamma_A/H_{\rm sb}\ll1$ and $\Gamma_A\tau_{\rm sb}\ll1$, with $H_{\rm sb}$ and $\tau_{\rm sb}$ denoting the Hubble scale and physical duration of symmetry breaking. The complementary kinetic regime is a finite-temperature gauge--Higgs transport problem rather than a coherent-quench matching problem.
}
{Throughout, $g$ in the dimensional equations denotes the physical gauge coupling $g_{\rm phys}$. The illustrative benchmark expresses the homogeneous response in units of $v_\Phi$ and $(g_{\rm phys}v_\Phi)^{-1}$, while $g_{\rm phys}$ remains an independent microscopic input.}

The purpose of this paper is to analyze that coherent transition problem in detail. Our main result is conceptually simple: the familiar adiabatic abundance relation is not the unique prediction of the post-inflationary mechanism, but one branch of an exact quench problem. By rewriting the homogeneous isotropic condensate in a canonical conformal-time variable, we derive a unified action variable that interpolates between the quartic and quadratic regimes. This construction makes it possible to identify the adiabatic reference branch, derive the exact quench-work and excess-energy equations, and define a coherent survival factor $f_{\rm coh}$ that multiplicatively renormalizes the final abundance,
\begin{equation}
\Omega_{Q,0}^{\rm quench}=f_{\rm coh}\,\Omega_{Q,0}^{\rm ad}.
\end{equation}
The standard adiabatic abundance contour is recovered only in the adiabatic limit $f_{\rm coh}\to 1$.

To sharpen the post-breaking dynamics beyond adiabatic matching, we next derive the homogeneous matrix formulation of the isotropic condensate, decompose the fluctuations under the diagonal $SO(3)$ preserved by the background, and obtain the full $1\oplus 3\oplus 5$ channel split. This exposes a soft traceless-symmetric quintet whose homogeneous quadratic gap is generated only by the symmetry-breaking mass. We show that this sector exhibits a $k=0$ vacuum obstruction in the unbroken phase, obeys a regulated ultraviolet adiabaticity bound, and is quartically stabilized rather than tachyonic. These results do not replace the full finite-$k$ helicity analysis, but they identify the infrared channel structure and selection rules that any complete treatment must accommodate.

The strategy of the paper is therefore to isolate the post-inflationary ingredients that can be brought under analytic and numerical control before the full gauge-Higgs transfer problem is attacked. Within that domain the treatment is closed: the coherent branch is renormalized by a calculable survival factor, the departure from the adiabatic benchmark is tracked by an excess-energy equation, and the soft homogeneous sector is organized by a definite channel hierarchy. This provides the natural matching framework for subsequent finite-$k$ and nonlinear extensions.

The paper is organized as follows. In \cref{sec:background} we review the isotropic post-inflationary condensate and derive the effective homogeneous Lagrangian. \Cref{sec:canonical} gives the canonical reduction and unified action variable, including the quartic-to-quadratic adiabatic coefficient. \Cref{sec:quenchwork} develops the quench-work, excess-energy, and coherent-abundance formulas. \Cref{sec:numerics} presents local and FRW numerical diagnostics. \Cref{sec:matrix} derives the homogeneous matrix formulation and diagonal-$SO(3)$ channel split, while \cref{sec:cubic} derives the cubic selection rules. \Cref{sec:soft} analyzes the soft quintet, including the zero-mode obstruction, regulated adiabatic bound, and quartic stabilization. \Cref{sec:pheno} develops the phenomenological interpretation and finite-$k$ consistency estimates, and \cref{sec:conclusions} summarizes the main lessons and immediate extensions. Technical derivations are collected in the appendices.

\section{Post-inflationary isotropic condensate and effective homogeneous dynamics}\label{sec:background}

\subsection{Isotropic SU(2) condensate after inflation}

We adopt the standard isotropic SU(2) background,
\begin{equation}
A_i^a(t)=a(t)Q(t)\,\delta_i^a,
\qquad A_0^a=0,
\label{eq:isotropic_ansatz}
\end{equation}
where $a(t)$ is the scale factor and $Q(t)$ is the gauge amplitude. This ansatz is consistent because spatial rotations can be compensated by SU(2) gauge rotations. The diagonal subgroup inherited from this locking is the organizing symmetry of the homogeneous fluctuation problem discussed later. The existence and attractor properties of the isotropic branch have been studied extensively in the axion--SU(2) literature \cite{adshead_wyman_chromonatural_2012,adshead_martinec_wyman_perturbations_2013,adshead_sfakianakis_wyman_higgsed_2016,wolfson_maleknejad_komatsu_attractive_2020,wolfson_etal_isotropic_2021}.

Once the axion driving term has become negligible after inflation, and after SU(2) acquires an effective mass $m(t)$ through spontaneous symmetry breaking, the homogeneous gauge amplitude obeys
\begin{equation}
\ddot Q+3H\dot Q+\left(\dot H+2H^2+m(t)^2\right)Q+2g^2Q^3=0,
\label{eq:Qeq_full}
\end{equation}
where $g$ is the SU(2) gauge coupling and $H=\dot a/a$ is the Hubble parameter. In the unbroken phase $m(t)=0$, this is a quartic oscillator dressed by expansion. In the broken phase the dynamics cross over to those of a massive vector. The inherited-condensate mechanism of \cite{fujita_murai_nakayama_yin_2024} analyzes this transition adiabatically and infers the final abundance from the preserved invariant. Our aim is to retain the same homogeneous starting point while treating the transition itself as a dynamical quench.

\subsection{Effective homogeneous Lagrangian}

A compact effective Lagrangian reproducing \eqref{eq:Qeq_full} is
\begin{equation}
L_{\rm eff}=\frac{3}{2}a^3\Big[(\dot Q+HQ)^2-g^2Q^4-m(t)^2Q^2\Big].
\label{eq:Leff_Q}
\end{equation}
The overall factor $3/2$ is part of the normalization inherited from the isotropic $SU(2)$ ansatz and is essential for the quartic and mass terms to reproduce the coefficients in \eqref{eq:Qeq_full}. Varying with respect to $Q$ gives
\begin{equation}
\begin{aligned}
0&=\frac{\dd}{\dd t}\left[3a^3(\dot Q+HQ)\right]-3a^3\left[H(\dot Q+HQ)-2g^2Q^3-m^2Q\right] \\
&=3a^3\Big[\ddot Q+3H\dot Q+(\dot H+2H^2+m^2)Q+2g^2Q^3\Big].
\end{aligned}
\end{equation}
which is precisely \eqref{eq:Qeq_full}. The utility of \eqref{eq:Leff_Q} is that it isolates the post-inflationary problem in a one-dimensional mechanical system whose time dependence enters only through the expansion and the mass profile. The rest of the paper exploits that reduction systematically.

\subsection{Local fast-oscillation limit}

When the oscillation timescale is much shorter than the Hubble timescale during the transition, the system can be treated locally as a nonlinear oscillator with slowly varying frequency. Neglecting expansion over a few oscillations gives
\begin{equation}
\ddot Q+\left[m(t)^2+2g^2Q^2\right]Q=0.
\label{eq:local_quench_eq}
\end{equation}
This local problem is not the complete cosmological evolution, but it is extremely useful because it exposes the quartic-to-quadratic matching problem in a clean form. In particular, it makes it possible to derive the adiabatic coefficient in closed form and to verify directly how fast quenches depart from adiabatic matching. We use the local problem in \cref{sec:numerics} as a high-precision diagnostic of endpoint matching, while the FRW calculation provides the cosmological consistency check.

\section{Canonical reduction and unified action variable}\label{sec:canonical}

\subsection{Canonical variable $X=aQ$}

The canonical simplification begins by passing to conformal time,
\begin{equation}
\dd \eta = \frac{\dd t}{a(t)},
\end{equation}
and defining the canonical variable
\begin{equation}
X(\eta)=a(\eta)Q(\eta).
\label{eq:Xdef}
\end{equation}
Since
\begin{equation}
\dot Q+HQ=\frac{X'}{a^2},
\end{equation}
where a prime denotes $\dd/\dd\eta$, the action becomes
\begin{equation}
S=\frac{3}{2}\int \dd \eta\,\left[X'^2-a^2m^2X^2-g^2X^4\right].
\end{equation}
Multiplying the full action by any nonzero constant leaves the Euler--Lagrange equation unchanged. Dividing by $3$ and adopting a canonical normalization therefore gives
\begin{equation}
\mathcal{L}_X=\frac{1}{2}X'^2-\frac{1}{2}\Omega(\eta)^2X^2-\frac{1}{2}g^2X^4,
\qquad
\Omega(\eta)=a(\eta)m(\eta),
\label{eq:LX}
\end{equation}
with Hamiltonian
\begin{equation}
\mathcal{H}_X=\frac{1}{2}P_X^2+\frac{1}{2}\Omega^2X^2+\frac{1}{2}g^2X^4.
\label{eq:HX}
\end{equation}
The equation of motion is therefore
\begin{equation}
X''+\left[\Omega(\eta)^2+2g^2X^2\right]X=0.
\label{eq:Xeq}
\end{equation}
This frictionless canonical form is the backbone of the entire analysis.

\subsection{Quartic and quadratic limits}

The two asymptotic regimes of \eqref{eq:Xeq} are immediate:
\begin{itemize}
\item In the quartic regime $\Omega^2\ll g^2X^2$, one has $X''+2g^2X^3\simeq 0$. The turning-point amplitude of $X$ is approximately constant, hence $Q=X/a\propto a^{-1}$.
\item In the quadratic regime $\Omega^2\gg g^2X^2$, one has $X''+\Omega^2X\simeq 0$. For constant late-time $m$, adiabatic invariance implies $A_X\propto \Omega^{-1/2}\propto a^{-1/2}$, hence $Q\propto a^{-3/2}$.
\end{itemize}
These are precisely the scalings familiar from the inherited-condensate mechanism. The present reformulation shows that both arise from a single canonical Hamiltonian.

\subsection{Unified action variable}

Let $A$ denote the turning-point amplitude of $X$. The energy at fixed $A$ and $\Omega$ is
\begin{equation}
E(A,\Omega)=\frac{1}{2}\Omega^2A^2+\frac{1}{2}g^2A^4.
\end{equation}
The unified action variable is
\begin{equation}
J(A,\Omega)=4\int_0^A \dd X\,
\sqrt{2\left[E(A,\Omega)-\frac{1}{2}\Omega^2X^2-\frac{1}{2}g^2X^4\right]}.
\label{eq:J_unified}
\end{equation}
This expression interpolates continuously between the quartic and quadratic regimes and is the appropriate invariant for adiabatic symmetry breaking.

In the quartic limit,
\begin{equation}
J_4=4gA^3\int_0^1 \dd u\,\sqrt{1-u^4}
\equiv C_4 gA^3,
\end{equation}
with
\begin{equation}
C_4=4\int_0^1 \dd u\,\sqrt{1-u^4}.
\end{equation}
The integral is elementary in Beta-function form. Setting $x=u^4$ gives
\begin{equation}
\int_0^1 \dd u\,\sqrt{1-u^4}=\frac14\int_0^1 \dd x\,x^{-3/4}(1-x)^{1/2}=\frac14 B\!\left(\frac14,\frac32\right),
\end{equation}
so that
\begin{equation}
C_4=B\!\left(\frac14,\frac32\right)=\frac{\Gamma(1/4)\Gamma(3/2)}{\Gamma(7/4)}=\frac{2\sqrt{\pi}\,\Gamma(1/4)}{3\Gamma(3/4)}.
\end{equation}
In the quadratic limit,
\begin{equation}
J_2=\pi \Omega A^2.
\end{equation}
Matching these limits under adiabatic evolution gives the quartic-to-quadratic coefficient
\begin{equation}
\Omega_f A_f^2=C_{\rm ad}\, gA_i^3,
\qquad
C_{\rm ad}=\frac{2\Gamma(1/4)}{3\sqrt{\pi}\Gamma(3/4)}\simeq 1.1128357889.
\label{eq:Cad}
\end{equation}
This sharpens the familiar parametric relation $mQ_{\rm aft}^2\sim gQ_{\rm bef}^3$ to an exact controlled coefficient.

\section{Quench work, excess energy, and coherent abundance renormalization}\label{sec:quenchwork}

\subsection{Energy injection by the time-dependent mass}

Because the Hamiltonian \eqref{eq:HX} depends on time only through $\Omega(\eta)$, its total derivative is simply
\begin{equation}
\frac{\dd H}{\dd\eta}=\frac{\partial H}{\partial \eta}=\Omega(\eta)\Omega'(\eta)X(\eta)^2.
\label{eq:energy_identity}
\end{equation}
This is the quench-work identity of the homogeneous canonical system. It states that the only source of energy injection into the homogeneous canonical system is the explicit time dependence of the symmetry-breaking mass profile.

\subsection{Adiabatic reference branch and excess energy}

Let $J_i$ be the early-time action of the quartic condensate. The adiabatic reference branch is defined by keeping this action fixed while allowing $\Omega(\eta)$ to vary,
\begin{equation}
E_{\rm ad}(\eta)=E\big(J_i,\Omega(\eta)\big).
\end{equation}
Using $J(E,\Omega)=\oint P\,\dd X$, one has the differential identities
\begin{equation}
\frac{\partial J}{\partial E}=T(E,\Omega),
\qquad
\frac{\partial J}{\partial \Omega}=-T(E,\Omega)\,\Omega\,\langle X^2\rangle_{E,\Omega},
\end{equation}
where $T$ is the oscillation period and $\langle X^2\rangle$ denotes the cycle average at fixed $(E,\Omega)$. Differentiating $J(E,\Omega)$ along the true trajectory and using \eqref{eq:energy_identity} gives
\begin{equation}
\frac{\dd J}{\dd\eta}=T\left[\frac{\dd E}{\dd\eta}-\Omega\Omega'\langle X^2\rangle\right].
\end{equation}
The oscillation average of the bracket vanishes, reproducing the usual adiabatic theorem in the present nonlinear setting.

Define the excess energy relative to the adiabatic branch,
\begin{equation}
\Delta E(\eta)=E(\eta)-E_{\rm ad}(\eta).
\end{equation}
Then
\begin{equation}
\Delta E' = \Omega\Omega'\Big[X^2-\langle X^2\rangle_{J_i,\Omega}\Big].
\label{eq:deltaEeq}
\end{equation}
This formula identifies the source of departure from the adiabatic branch: the mismatch between the instantaneous trajectory and the cycle average required by fixed action.

\subsection{Coherent abundance closure}

In the late quadratic regime,
\begin{equation}
E\simeq \frac{\Omega J}{2\pi},
\qquad
J\simeq \pi \Omega A^2.
\end{equation}
The oscillation-averaged energy density of the massive vector condensate is
\begin{equation}
\langle \rho_Q\rangle=\frac{3}{2}m^2Q_{\rm amp}^2,
\end{equation}
so the comoving coherent number density is
\begin{equation}
 n_{\rm coh}a^3 = \frac{\langle\rho_Q\rangle a^3}{m} = \frac{3}{2}ma^3Q_{\rm amp}^2 = \frac{3}{2\pi}J_{\rm late}.
\label{eq:ncohJ}
\end{equation}
Therefore the final coherent abundance is proportional to the late-time action. This suggests the natural survival factor
\begin{equation}
 f_{\rm coh}\equiv \frac{J_{\rm late}}{J_{\rm early}}.
\label{eq:fcohdef}
\end{equation}
The coherent abundance in the general quench problem is then
\begin{equation}
\Omega_{Q,0}^{\rm quench}=f_{\rm coh}\,\Omega_{Q,0}^{\rm ad}.
\label{eq:Omegaquench}
\end{equation}
This is the central replacement rule of the paper.

In particular, the benchmark adiabatic abundance contour is the zero-excess-energy branch of the quench problem. If the inherited-condensate abundance scales as
\begin{equation}
\Omega_{Q,0}^{\rm ad}\propto m\,g^{-1/2}c^{-3/4},
\end{equation}
with $c$ denoting the model-dependent inflationary parameter defined in ref.~\cite{fujita_murai_nakayama_yin_2024}, the corrected observed-abundance condition becomes
\begin{equation}
 f_{\rm coh}\,m\,g^{-1/2}c^{-3/4}=\text{const.}
\label{eq:corrected_line}
\end{equation}
At fixed $(g,c)$ the required mass shifts by $m\to m/f_{\rm coh}$, while at fixed $(m,c)$ the required gauge coupling shifts by $g\to g f_{\rm coh}^2$.

\section{Numerical quench diagnostics for the coherent sector}\label{sec:numerics}

This section summarizes the numerical checks used to validate the coherent-sector formulas and to illustrate the domain in which adiabatic matching remains reliable. The local and FRW evolutions are solved with adaptive Runge--Kutta integration. Late-time amplitudes or actions are extracted from the average of the last turning points after the evolution has entered its asymptotic regime. The numerical procedure is described in a self-contained way in appendix~\ref{app:numerics}.

\subsection{Local quench: convergence to the adiabatic coefficient}

\begin{figure}[t]
  \centering
  \includegraphics[width=\columnwidth]{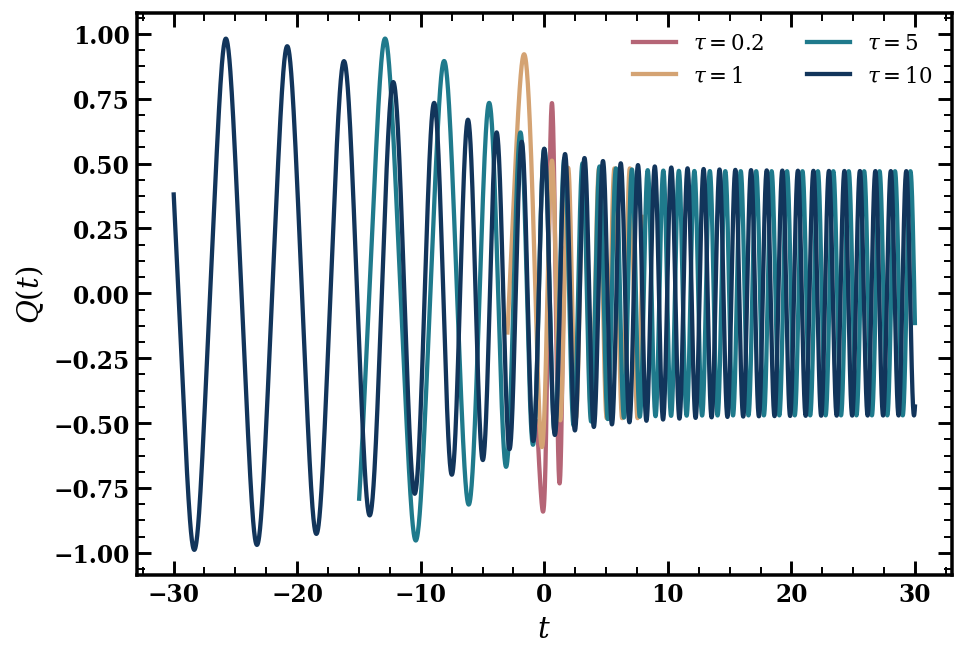}
  \caption{Representative local quartic-to-quadratic quench trajectories for the homogeneous oscillator \eqref{eq:local_quench_eq}. The quench width $\tau$ controls the departure from adiabatic endpoint matching.}
  \label{fig:localtraj}
\end{figure}

\begin{figure}[t]
  \centering
  \includegraphics[width=\columnwidth]{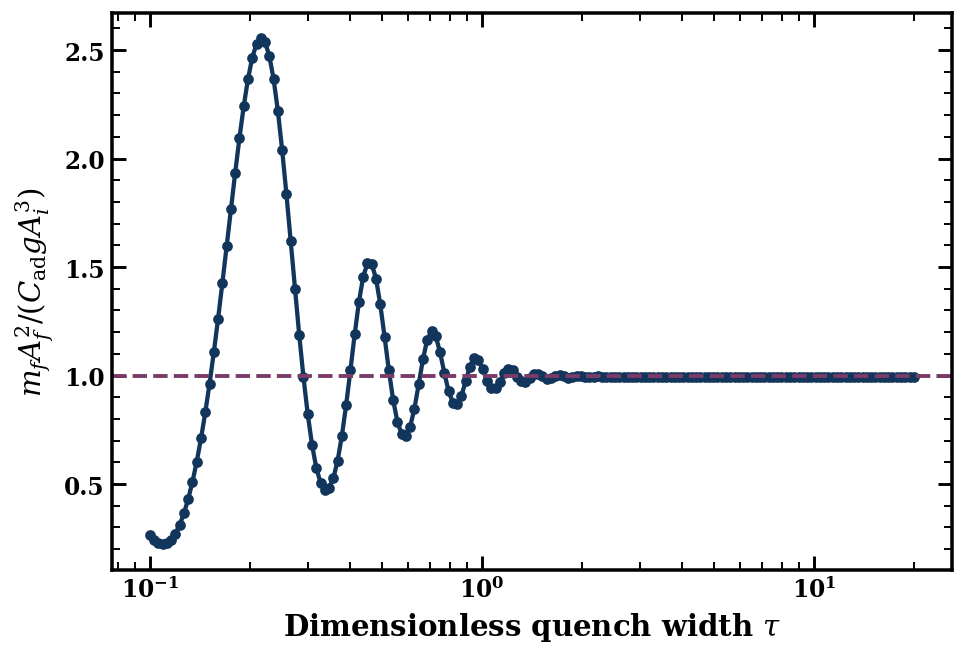}
  \caption{Convergence of the local-quench endpoint to the analytic adiabatic coefficient $C_{\rm ad}$ in \eqref{eq:Cad}. Slow smooth quenches approach the analytic coefficient, while rapid quenches deviate at order unity.}
  \label{fig:localconv}
\end{figure}

We begin by solving the local equation \eqref{eq:local_quench_eq} with a smooth monotonic mass turn-on,
\begin{equation}
 m(t)^2=m_f^2\frac{1+\tanh(t/\tau)}{2},
\end{equation}
for representative parameters $(g,m_f,A_i)=(1,5,1)$. The hyperbolic-tangent profile is used here as an effective interpolation characterized only by an asymptotic mass and a transition width; in a microphysical Higgs completion these parameters would be inherited from $m(\eta)=gv(\eta)$ after matching the scalar evolution to a smooth crossover. The present scan is therefore best read as a response map of the coherent sector to the transition width and asymptotic mass, rather than as a first-principles Higgs-sector survey. The late-time amplitude $A_f$ is extracted from the oscillation envelope and compared with the analytic adiabatic coefficient \eqref{eq:Cad}. The results are displayed in \cref{fig:localtraj,fig:localconv}. Rapid quenches clearly depart from the adiabatic branch, while slow transitions converge to it at the sub-percent level. In the representative scan the ratio $(m_fA_f^2)/(C_{\rm ad}gA_i^3)$ equals $1.304$ for $\tau=0.2$, $0.938$ for $\tau=1$, and $0.9945$ for $\tau=5$ and $20$.

\subsection{Illustrative microphysical interpretation of the quench width}

The width parameter $\tau$ in the effective interpolation $m(\eta)=m_f[1+\tanh((\eta-\eta_\mathrm{sb})/\tau)]/2$ should be interpreted as an approximation for the timescale on which the order parameter of the symmetry-breaking sector approaches its late-time expectation value. In a simple Higgs realization with an adjoint or fundamental scalar field $\Phi$ and potential
\begin{equation}
V(\Phi)=\frac{\lambda_\Phi}{4}\left(|\Phi|^2-v_\Phi^2\right)^2,
\end{equation}
the effective gauge-boson mass is $m(\eta)=g\,v(\eta)$ with $v(\eta)=\sqrt{2\langle|\Phi|^2\rangle}$. Linearization around the broken minimum gives the scalar oscillation scale
\begin{equation}
m_\Phi^2\simeq 2\lambda_\Phi v_\Phi^2,
\end{equation}
so a smooth classical crossover is expected to occur over a conformal-time interval of order $\tau_\Phi\sim (a_\mathrm{sb}m_\Phi)^{-1}$, up to order-unity corrections from Hubble damping and from the details of the scalar initial conditions \cite{sato_gravitational_2022,nakayama_yin_symmetry_2021,salehian_vector_2021,adshead_sfakianakis_wyman_higgsed_2016}. In this language the dimensionless canonical scan probes the ratio of the scalar relaxation time to the intrinsic oscillation time of the inherited condensate. The slow-quench regime corresponds to $a_\mathrm{sb}m_\Phi$ well below the characteristic quartic oscillation frequency of the condensate, while the rapid-quench regime corresponds to comparable or larger scalar-curvature scales.

This mapping is intentionally illustrative rather than exhaustive. The purpose of the scan is not to claim a universal Higgs prediction from the tanh profile, but to identify how the coherent branch responds once the symmetry-breaking timescale becomes comparable to the oscillation timescale. The numerical results, therefore, isolate the dynamical sensitivity of the inherited condensate to the transition rate, which is the part of the problem that is independent of the detailed ultraviolet completion.

\subsection{{An illustrative thermal order-parameter benchmark}{An illustrative thermal order-parameter benchmark}}

{The effective interpretation above can be made more explicit with a minimal thermal background for the symmetry-breaking order parameter. We consider a real scalar $\phi$ with finite-temperature effective potential}
\begin{equation}
V_{\rm eff}(\phi,\eta)=\frac{\lambda_\Phi}{4}(\phi^2-v_\Phi^2)^2+\frac{c_T}{2}T(\eta)^2\phi^2,
\qquad
T(\eta)=\frac{T_0}{a(\eta)}.
\label{eq:Veffthermal}
\end{equation}
The homogeneous scalar obeys the following condition
\begin{equation}
\phi''+2\frac{a'}{a}\phi'+a^2\Big[\lambda_\Phi\phi(\phi^2-v_\Phi^2)+c_TT(\eta)^2\phi\Big]=0,
\label{eq:phieomthermal}
\end{equation}
while the inherited vector mass is determined self-consistently by
\begin{equation}
m(\eta)=g\,\phi(\eta).
\label{eq:micro_mass}
\end{equation}
The critical temperature is $T_c=\sqrt{\lambda_\Phi/c_T}\,v_\Phi$, and the transition width is no longer a free label but an output of the scalar roll.

{
To separate the dimensionless response from the microscopic coupling, we write $g\equiv g_{\rm phys}$ and define
\begin{equation*}
\begin{aligned}
\hat\eta&=g_{\rm phys}v_\Phi\eta,
&\hat X&=X/v_\Phi,\\
\hat\phi&=\phi/v_\Phi,
&\hat T&=T/v_\Phi,\\
\hat\lambda_\Phi&=\lambda_\Phi/g_{\rm phys}^2,
&\hat c_T&=c_T/g_{\rm phys}^2.
\end{aligned}
\end{equation*}
The coupled homogeneous equations then take the dimensionless form
\begin{align*}
\hat X_{,\hat\eta\hat\eta}+\left(a^2\hat\phi^2+2\hat X^2\right)\hat X&=0,\\
\hat\phi_{,\hat\eta\hat\eta}+2\frac{a_{,\hat\eta}}{a}\hat\phi_{,\hat\eta}
+a^2\left[\hat\lambda_\Phi\hat\phi(\hat\phi^2-1)+\hat c_T\hat T^2\hat\phi\right]&=0.
\end{align*}
Thus the plotted benchmark specifies the hatted ratios and uses $(g_{\rm phys}v_\Phi)^{-1}$ as the conformal-time unit, while microscopic embeddings retain $g_{\rm phys}$ as an independent parameter. Hats are suppressed below only in the numerical labels.
}

{This benchmark is a thermal order-parameter background for the symmetry-breaking sector. The inherited $SU(2)$ condensate is treated in the kinetically resolved coherent regime. The consistency conditions are
\begin{equation}
\frac{\Gamma_A}{H_{\rm sb}}\ll1,
\qquad
\Gamma_A\tau_{\rm sb}\ll1,
\label{eq:kinetic_regime}
\end{equation}
where $\Gamma_A$ denotes the relevant gauge-boson thermalization or decoherence rate, $H_{\rm sb}$ is the Hubble scale at symmetry breaking, and $\tau_{\rm sb}$ is the physical transition time. {Equation~\eqref{eq:kinetic_regime} constrains the physical coupling $g_{\rm phys}$ independently of the normalization above. In a populated weakly coupled non-Abelian plasma, color-randomization or damping scales enter parametrically at order $g_{\rm phys}^2T\ln(1/g_{\rm phys})$, while kinetic equilibration rates enter at order $g_{\rm phys}^4T\ln(1/g_{\rm phys})$, with the precise rate determined by the process and occupancy \cite{ArnoldSonYaffe1999,ArnoldMooreYaffe2003}. At $g_{\rm phys}\sim1$, these rates carry order-unity coupling factors in an ordinary populated thermal gauge plasma. Coherent protection therefore arises in sufficiently weakly coupled, dilute, or kinetically secluded gauge sectors. The present benchmark represents this kinetically resolved class, with $g_{\rm phys}$ chosen to satisfy \eqref{eq:kinetic_regime}. The thermal mass of the order parameter may be dominated by scalar self-interactions or by a secluded bath, while the inherited gauge zero mode remains sufficiently weakly coupled, dilute, or kinetically decoupled during the crossover. The complementary regime $\Gamma_A\gtrsim H_{\rm sb}$ or $\Gamma_A\tau_{\rm sb}\gtrsim1$ is described by finite-temperature gauge--Higgs transport.}
}

\begin{figure*}[t]
  \centering  \includegraphics[width=0.92\textwidth]{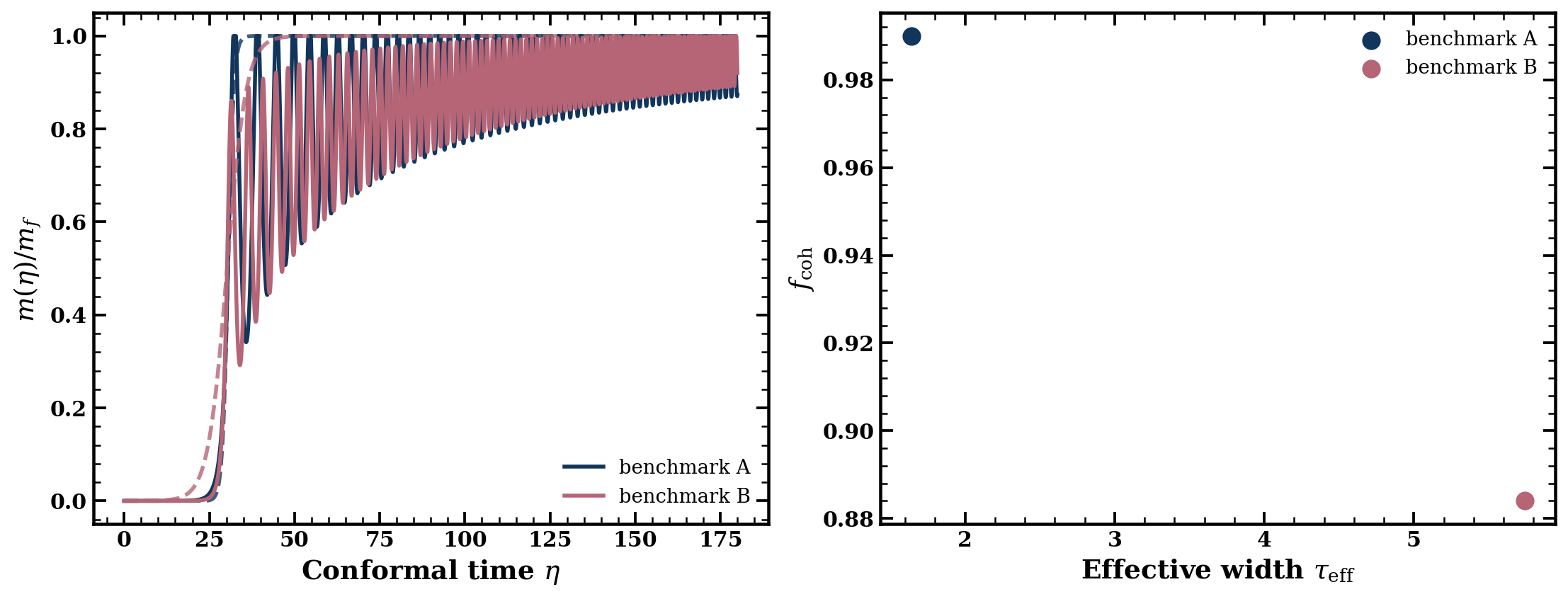}
  \caption{{Illustrative dimensionless thermal order-parameter benchmarks defined by \eqref{eq:Veffthermal}--\eqref{eq:micro_mass}. Fields are expressed in units of $v_\Phi$, and conformal time in units of $(g_{\rm phys}v_\Phi)^{-1}$; the plotted parameters are $\hat\lambda_\Phi=\lambda_\Phi/g_{\rm phys}^2$ and $\hat c_T=c_T/g_{\rm phys}^2$, with hats suppressed in the panel labels. Left: normalized order-parameter profiles and matched tanh fits used to define $\hat\tau_{\rm eff}=g_{\rm phys}v_\Phi\tau_{\rm eff}$. Right: the corresponding values of $(\hat\tau_{\rm eff},f_{\rm coh})$. The detailed symmetry-breaking profile supplies information beyond the effective width and thereby influences the coherent survival factor.}}
  \label{fig:microbench}
\end{figure*}

To compare with the canonical scan we define an effective width $\tau_{\rm eff}$ from the $10\%$--$90\%$ rise time of $m(\eta)/m_f$, and then solve the coherent quench problem using the full benchmark profile rather than an analytic interpolation. Two representative thermal benchmarks are shown in \cref{fig:microbench}: a near-adiabatic case with $(\lambda_\Phi,T_0/T_c)=(0.5,1.05)$ and a more rapidly evolving case with $(\lambda_\Phi,T_0/T_c)=(2.0,1.15)$, {both for $\hat c_T=1$, using $v_\Phi$ as the field unit and $g_{\rm phys}v_\Phi$ as the inverse-time unit; hats are suppressed in the figure labels.} They yield
\begin{equation}
(\tau_{\rm eff},f_{\rm coh})\simeq (1.64,0.990),
\qquad
(5.74,0.884),
\label{eq:benchpairs}
\end{equation}
respectively, with {$a_{\rm sb}m_\Phi/(g_{\rm phys}v_\Phi)\simeq 1.49$ and $2.87$. The quoted points specify dimensionless response curves. A microscopic realization chooses $g_{\rm phys}$ together with $\lambda_\Phi=g_{\rm phys}^2\hat\lambda_\Phi$ and $c_T=g_{\rm phys}^2\hat c_T$, within the kinetically resolved domain defined by \eqref{eq:kinetic_regime}. The curves therefore represent the family of microscopic embeddings associated with these hatted ratios.} The figure illustrates two points that are phenomenologically relevant. Realistic symmetry-breaking backgrounds do populate the same order-unity range of coherent suppression found in the effective scan. At the same time, the full shape of the scalar profile matters in addition to any single width parameter; the benchmark points do not collapse to a one-parameter family in $\tau_{\rm eff}$ alone. The model-agnostic scan and the explicit scalar benchmark are therefore complementary rather than redundant.

{The quench regime extends beyond this thermal illustration. It is expected whenever $|\Omega'/\Omega^2|\gtrsim1$ during the transition, equivalently when the symmetry-breaking time is comparable to the intrinsic oscillation period of the inherited condensate. This situation can arise in nonthermal Higgs relaxation after inflation, tachyonic or supercooled hidden-sector transitions, reheating-induced symmetry restoration followed by delayed breaking, modulus- or inflaton-controlled vector masses, and other late-forming dark-sector scenarios. In all such cases the adiabatic abundance relation is the slow-transition limit of the more general quench problem.
}

\subsection{FRW action conservation and excess energy}

\begin{figure}[t]
  \centering  \includegraphics[width=\columnwidth]{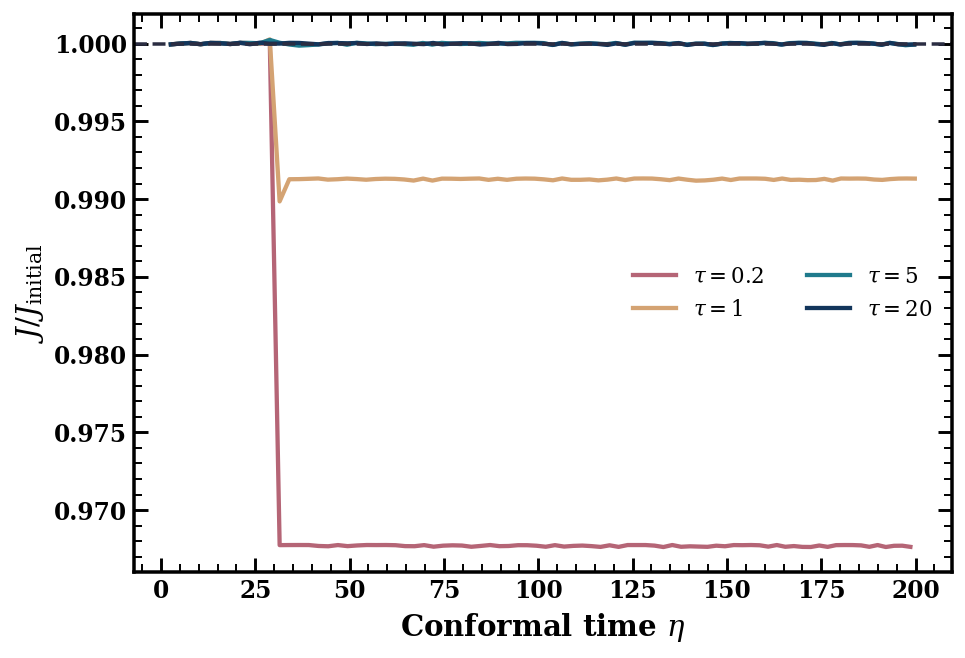}
  \caption{FRW action conservation test for the canonical quench problem. The action is preserved to excellent accuracy for slow transitions and shows a controlled deficit for rapid quenches.}
  \label{fig:frwaction}
\end{figure}

We next solve the full canonical FRW equation \eqref{eq:Xeq} in a radiation-era background with the same tanh mass profile. \Cref{fig:frwaction} shows the evolution of $J/J_{\rm initial}$ evaluated at turning points. The action is preserved to better than $10^{-5}$ for sufficiently slow quenches, while rapid transitions produce a visible drift. For $\tau=(0.2,1,5,20)$ we find relative drifts $(J_f-J_i)/J_i\simeq(-3.23\times 10^{-2},-8.70\times 10^{-3},-1.17\times 10^{-5},-1.48\times 10^{-5})$.

\begin{figure}[t]
  \centering
  \includegraphics[width=\columnwidth]{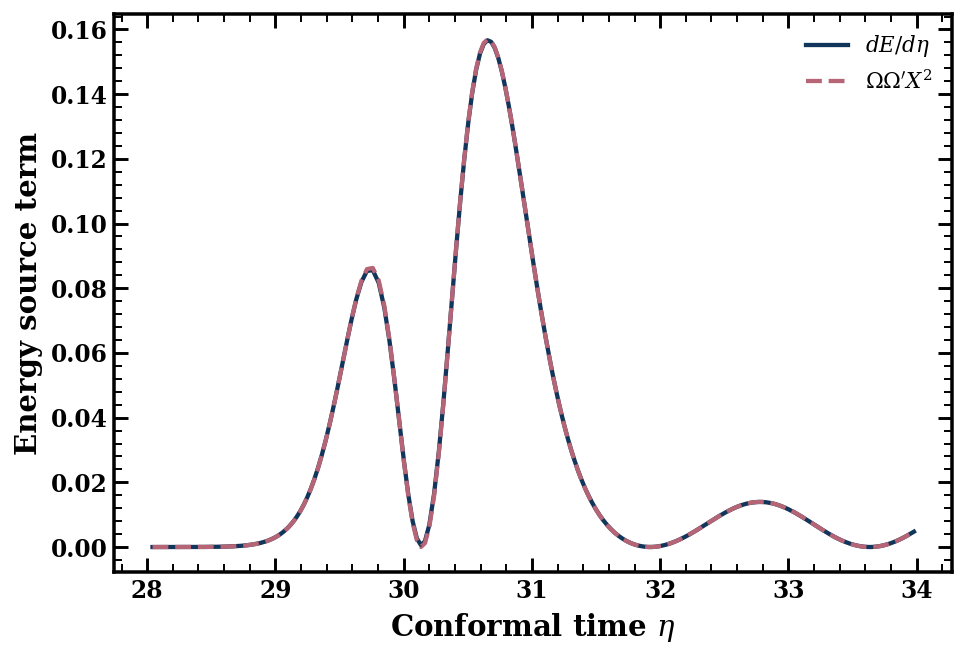}
  \caption{Numerical verification of the quench-work identity \eqref{eq:energy_identity} in a representative FRW run.}
  \label{fig:energyidentity}
\end{figure}

\begin{figure}[t]
  \centering
  \includegraphics[width=\columnwidth]{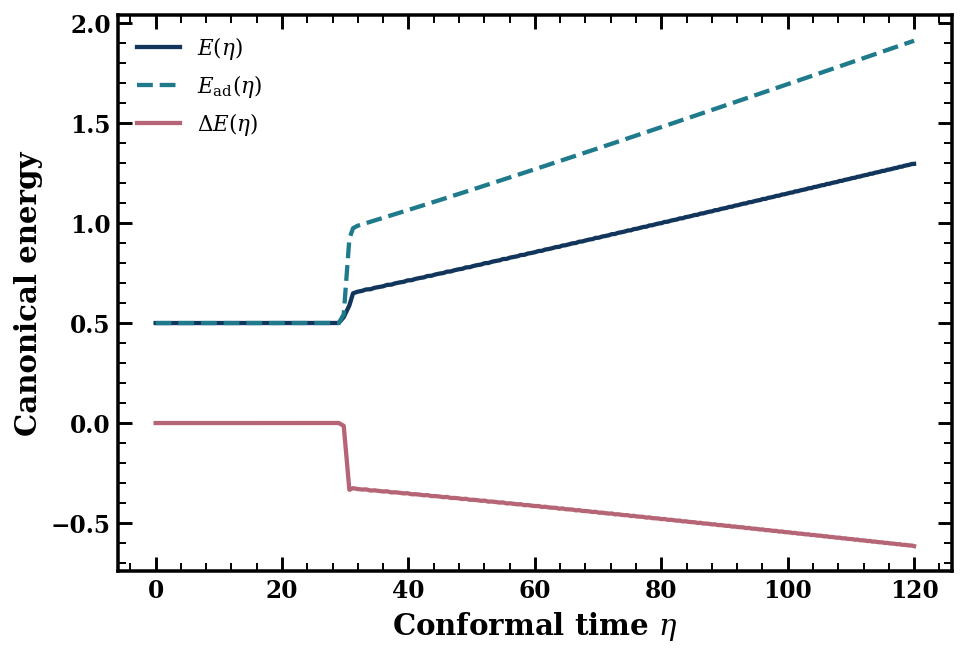}
  \caption{Energy along the true trajectory, the adiabatic reference branch, and the excess energy $\Delta E$. The benchmark adiabatic abundance contour corresponds to the zero-excess-energy branch.}
  \label{fig:excessenergy}
\end{figure}

\Cref{fig:energyidentity} compares the numerically differentiated Hamiltonian with the quench-work identity \eqref{eq:energy_identity}, while \cref{fig:excessenergy} shows the true energy, the adiabatic reference branch, and the excess energy for a representative run. In the benchmark case $(g,m_f,\tau)=(1,1,0.5)$ we obtain
\begin{equation}
J_i=3.49590019,
\qquad
J_f=2.37445908,
\qquad
f_{\rm coh}=0.67921249,
\end{equation}
with a transition-window relative RMS discrepancy of $3.17\times 10^{-3}$ in the energy-identity check. The final excess energy extracted directly from the adiabatic branch is $\Delta E_f^{\rm exact}=-1.06875602$, while the late-quadratic closure formula gives $\Delta E_f^{\rm quad}=-1.06168932$, confirming the consistency of the coherent-sector closure.

\subsection{Survival-factor scan and abundance shifts}

\begin{figure}[t]
  \centering
  \includegraphics[width=\columnwidth]{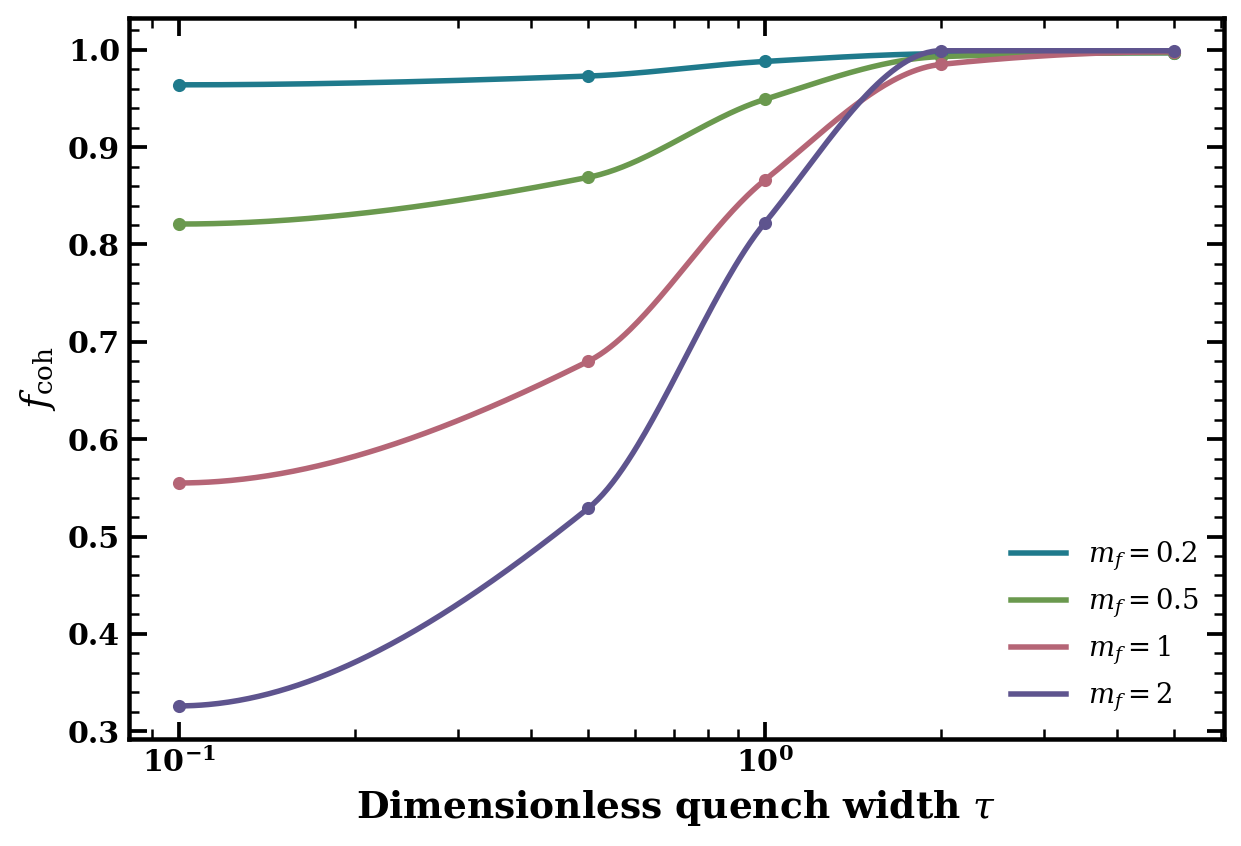}
  \caption{Representative scan of the coherent survival factor $f_{\rm coh}$. Rapid quenches and larger final masses suppress the coherent branch more strongly.}
  \label{fig:fcohscan}
\end{figure}

\begin{figure}[t]
  \centering
  \includegraphics[width=\columnwidth]{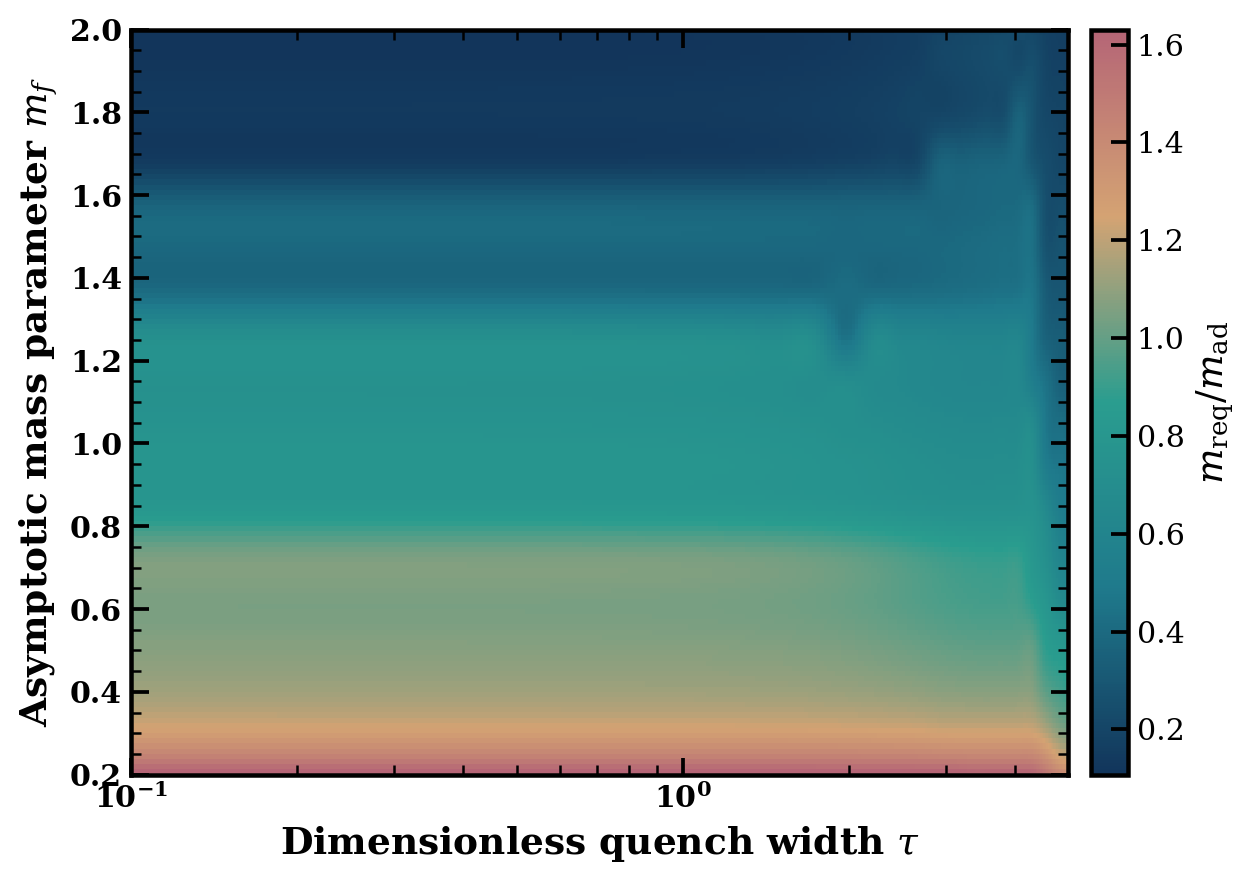}
  \caption{Illustrative abundance-shift heatmap based on the representative survival-factor scan. The plotted quantity is the mass-shift factor at fixed $(g,c)$ in \eqref{eq:corrected_line}. Both axes are dimensionless parameters of the canonical scan. The horizontal axis is the asymptotic mass parameter $m_f$ entering the canonical post-inflationary oscillator, measured in units of the initial quartic oscillation scale; the vertical axis is the corresponding dimensionless transition width $\tau$. The color bar gives the linear mass-shift factor at fixed $(g,c)$.}
  \label{fig:heatmap}
\end{figure}

A representative scan of $f_{\rm coh}$ is shown in \cref{fig:fcohscan}. The most robust interpretation of the scan is qualitative: the coherent abundance is only weakly renormalized for sufficiently slow quenches, but can be depleted at order unity once the transition is fast on the oscillation timescale. The resulting shift of the adiabatic abundance line is summarized in \cref{fig:heatmap}. Depending on $(m_f,\tau)$, the mass required at fixed $(g,c)$ can shift by factors ranging from a few percent to $\cO(3)$, while the effective direct-detection normalization of the dark-photon component inherits the same suppression. These shifts isolate a genuine dynamical correction associated with the symmetry-breaking stage; in concrete inflationary realizations they complement, rather than replace, the uncertainties tied to reheating and to the parameter $c$ controlling the inherited condensate amplitude \cite{fujita_murai_nakayama_yin_2024}.

\begin{figure}[t]
  \centering
  \includegraphics[width=\columnwidth]{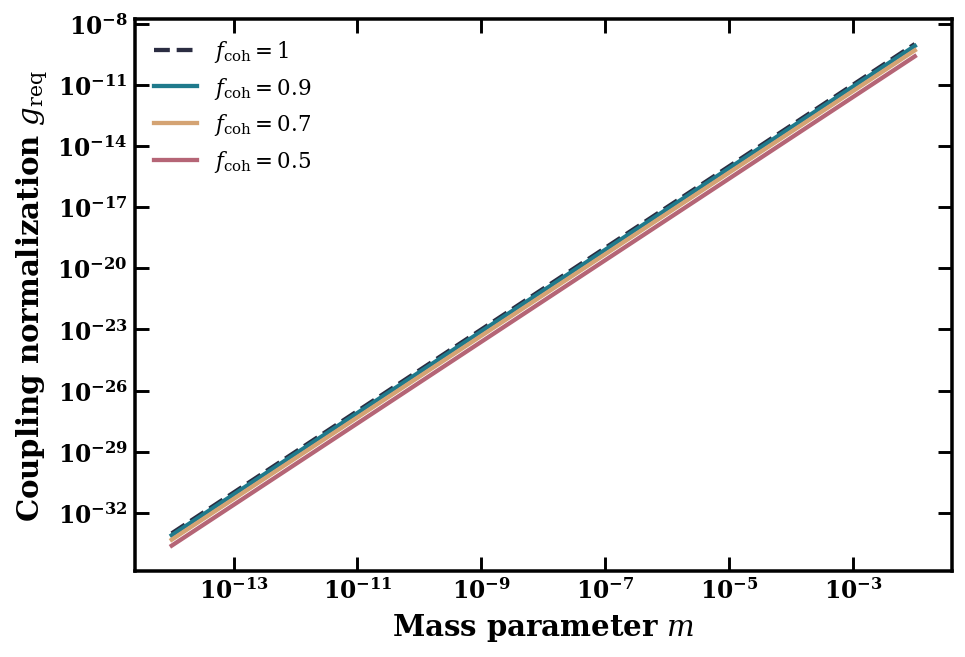}
  \caption{Schematic deformation of the coherent matching branch induced by non-adiabatic symmetry breaking. The dashed curve denotes the adiabatic branch $f_{\rm coh}=1$, while the solid curves show representative deformations generated by non-adiabatic symmetry-breaking quenches. The figure is intended as a model-agnostic summary of how the coherent branch moves once $f_{\rm coh}\neq1$, not as a substitute for a full microscopic parameter scan.}
  \label{fig:cohshift}
\end{figure}

\section{Homogeneous matrix formulation and diagonal-SO(3) channel decomposition}\label{sec:matrix}

\subsection{Matrix formulation}

To expose the structure of homogeneous fluctuations, it is convenient to treat the spatial gauge field as a general real $3\times 3$ matrix $X_i{}^a(\eta)$. In conformal time the homogeneous Lagrangian can be defined as:
\begin{equation}
L=\frac{1}{2}\Tr(X'^TX')-\frac{1}{2}\Omega^2\Tr(XX^T)-\frac{g^2}{4}\left[(\Tr XX^T)^2-\Tr\left((XX^T)^2\right)\right].
\label{eq:matrixL}
\end{equation}
For the isotropic background $X=q(\eta)\,\mathbb{1}$ this reduces to the single-field canonical Lagrangian \eqref{eq:LX}.

\subsection{Diagonal-SO(3) decomposition}

The isotropic background identifies gauge and spatial indices, leaving a diagonal $SO(3)$ unbroken. Any homogeneous fluctuation can therefore be decomposed into irreducible pieces under this diagonal group,
\begin{equation}
\delta X = \chi\,\mathbb{1}+A+T,
\label{eq:decomp}
\end{equation}
with
\begin{equation}
A^T=-A,
\qquad
T^T=T,
\qquad
\Tr T=0.
\end{equation}
This gives the full $1\oplus 3\oplus 5$ split of the homogeneous sector.

On expanding \eqref{eq:matrixL} to quadratic order around $X=q\,\mathbb{1}$ yields
\begin{equation}
\begin{aligned}
L_2={}&\frac{3}{2}\chi'^2+\frac{1}{2}\Tr(A'^TA')+\frac{1}{2}\Tr(T'^TT') \\
&-\frac{3}{2}(\Omega^2+6g^2q^2)\chi^2-\frac{1}{2}(\Omega^2+2g^2q^2)\Tr(AA^T)\\&-\frac{1}{2}\Omega^2\Tr(T^2).
\end{aligned}
\label{eq:L2channels}
\end{equation}
The homogeneous channel frequencies are expressed as:
\begin{equation}
\omega_{\rm tr}^2=\Omega^2+6g^2q^2,
\qquad
\omega_A^2=\Omega^2+2g^2q^2,
\qquad
\omega_T^2=\Omega^2.
\label{eq:omegachannels}
\end{equation}

Two features stand out. The trace and antisymmetric channels remain gapped even before symmetry breaking, because their quadratic terms are supplied by the quartic non-Abelian interaction. By contrast, the traceless-symmetric quintet is unique: its quadratic gap is generated only by the symmetry-breaking mass, and in the unbroken phase it is therefore gapless at $k=0$.

\begin{figure}[t]
  \centering
  \includegraphics[width=\columnwidth]{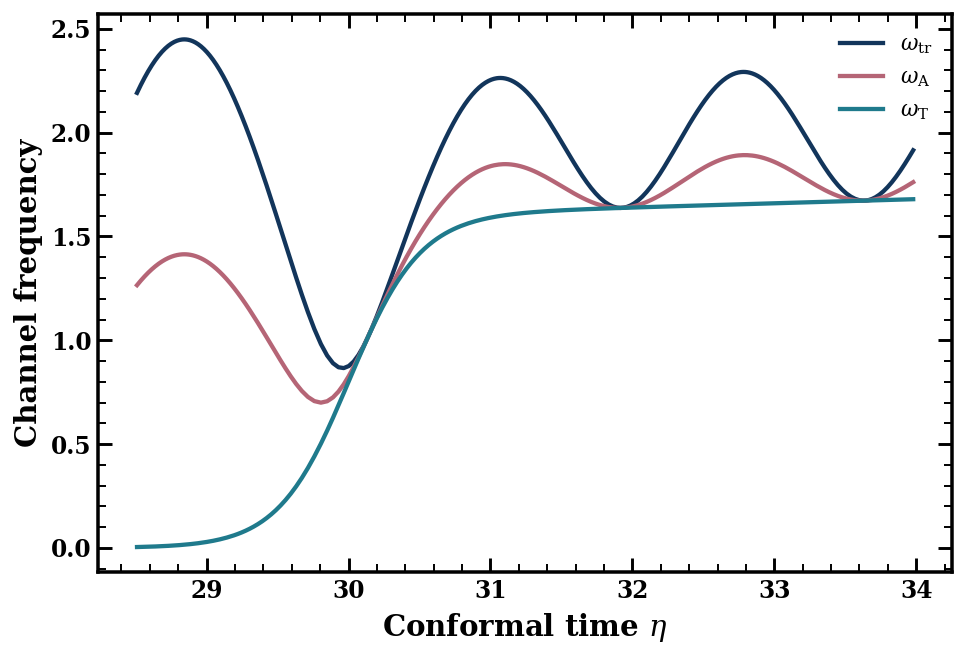}
  \caption{Homogeneous diagonal-$SO(3)$ channel frequencies \eqref{eq:omegachannels} in a representative quench. The traceless-symmetric quintet is the softest channel because its quadratic gap is generated only by the symmetry-breaking mass.}
  \label{fig:chfreq}
\end{figure}

\begin{figure}[t]
  \centering
  \includegraphics[width=\columnwidth]{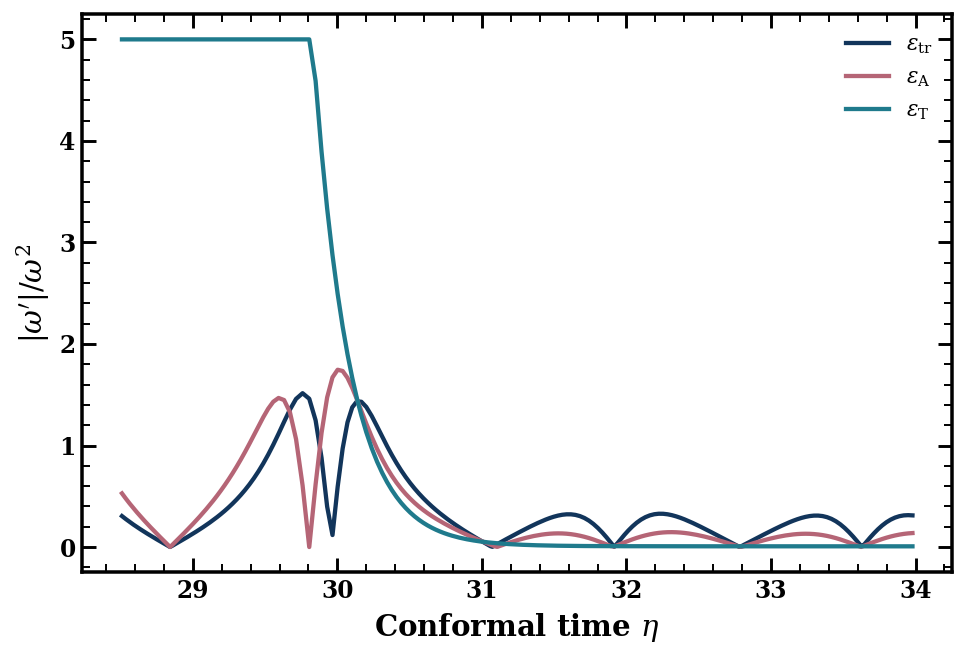}
  \caption{Local adiabaticity indicators for the homogeneous channels. The traceless-symmetric quintet is the least adiabatic because its frequency becomes parametrically small during the quench.}
  \label{fig:chadiab}
\end{figure}

\Cref{fig:chfreq,fig:chadiab} show the corresponding homogeneous frequency hierarchy and local adiabaticity indicators for a representative quench. In the benchmark run the minimum transition-window frequencies are approximately $(\omega_{\rm tr},\omega_A,\omega_T)=(0.866,0.700,6.19\times10^{-4})$, while the peak local adiabaticity indicators are $(1.54,2.66,6.52\times10^3)$. The main lesson is structural rather than numerical: the quintet is the unique homogeneous channel whose softness is directly tied to the symmetry-breaking quench.

\section{Cubic interaction structure}\label{sec:cubic}

The homogeneous channel ordering by itself does not determine the nonlinear energy-transfer network. To expose the leading nontrivial couplings we expand the potential to cubic order around $X=q\,\mathbb{1}$. The result is
\begin{equation}
V_3 = g^2 q\Big[6\chi^3+2\chi\Tr(AA^T)-\Tr(T^3)-\Tr(TAA^T)\Big].
\label{eq:V3}
\end{equation}
The algebraic identities entering this expression are reproduced explicitly in appendix~\ref{app:cubic} and appendix~\ref{app:quartic}.

Several aspects of this structure deserve emphasis. There is no $\chi\Tr(T^2)$ cubic coupling, so the trace mode does not couple directly to a pair of quintet excitations at leading nonlinear order. The quintet instead couples cubically only through self-interactions and through its interaction with the antisymmetric triplet. The interaction graph is therefore selective rather than democratic: softness of the quintet at quadratic order does not imply that the isotropic condensate immediately drains into every fluctuation channel at the first nonlinear step.

A simple dimensional estimate shows why this selectivity matters. The cubic vertices scale as $g^2 q$ times a fluctuation bilinear, so the associated local channel-transfer rates are parametrically of order $\Gamma_{\rm cubic}\sim g^2 q^2/\omega_{\rm gap}$ when a gapped daughter mode of frequency $\omega_{\rm gap}$ is kinematically available. The absence of a direct $\chi TT$ bridge therefore removes the most immediate coherent-to-quintet conversion path at cubic order and supports the bottleneck interpretation of the soft sector, even though a complete finite-$k$ calculation is still required for quantitative branching fractions.

This observation is phenomenologically important because it constrains the interpretation of coherent depletion. The cubic structure shows that the post-symmetry-breaking sector is subtle for two independent reasons: the quintet is soft at quadratic order, but it is also selectively coupled. Any claim about the detailed flow of energy into inhomogeneous sectors must therefore go beyond the homogeneous quadratic and cubic analysis performed here.

\section{Soft quintet: vacuum obstruction, regulated adiabatic bound, and quartic stabilization}\label{sec:soft}

\subsection{Zero-mode obstruction}

The homogeneous quintet frequency is simply
\begin{equation}
\omega_T^2(\eta,k=0)=\Omega(\eta)^2.
\end{equation}
Before symmetry breaking, $\Omega(\eta)\to 0$ and therefore $\omega_T(\eta_i,0)=0$. The homogeneous quintet thus does not admit a standard Gaussian adiabatic in-vacuum at $k=0$ in the unbroken phase. This is a direct statement of the homogeneous quadratic theory in temporal gauge. It should not be confused with the full gauge-fixed propagating finite-$k$ spectrum; rather, it characterizes the quadratic form of the reduced homogeneous sector inherited from the isotropic background. By contrast, the trace and antisymmetric channels remain gapped in the unbroken phase because of \eqref{eq:omegachannels}.

The significance of this result is conceptual. The problem in the soft sector is not merely that the quintet is produced more efficiently under some model-dependent prescription. The more basic point is that the unbroken-phase homogeneous limit does not support the standard quadratic vacuum one would use in an ordinary massive mode analysis. Any complete finite-$k$ production analysis must therefore regulate this IR subtlety explicitly, either by finite momentum or by finite volume.

\subsection{Regulated ultraviolet adiabatic bound}

\begin{figure}[t]
  \centering
  \includegraphics[width=\columnwidth]{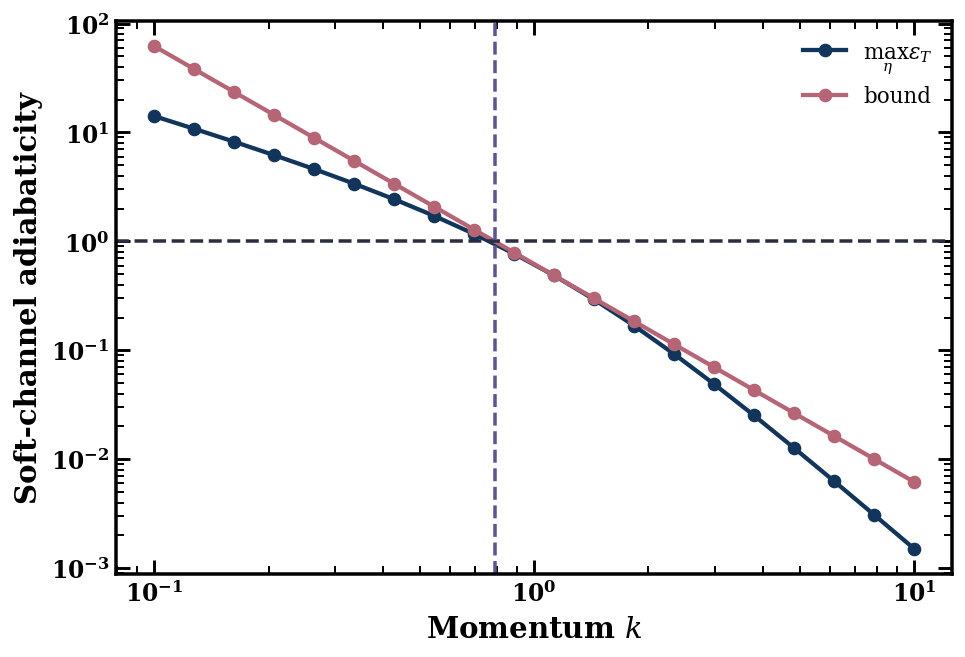}
  \caption{Regulated soft-channel adiabaticity bound. The shaded physical lesson is that nonadiabaticity is confined to an infrared band whose width is set by the quench rate.}
  \label{fig:softbound}
\end{figure}

For the minimal regulated continuation
\begin{equation}
\omega_T^2(\eta,k)=k^2+\Omega(\eta)^2,
\end{equation}
the local adiabaticity parameter is
\begin{equation}
\epsilon_T(\eta,k)=\frac{|\Omega\Omega'|}{(k^2+\Omega^2)^{3/2}}.
\label{eq:epssoft}
\end{equation}
The function $f(\Omega)=\Omega/(k^2+\Omega^2)^{3/2}$ is maximized at $\Omega=k/\sqrt{2}$, which implies the rigorous bound
\begin{equation}
\epsilon_T(\eta,k)\le \frac{2}{3\sqrt{3}}\frac{\sup_\eta |\Omega'(\eta)|}{k^2}.
\label{eq:softbound}
\end{equation}
Hence, for any target adiabaticity threshold $\epsilon_*$, there is a guaranteed adiabatic UV band
\begin{equation}
 k > k_{\rm ad}(\epsilon_*)\equiv \sqrt{\frac{2}{3\sqrt{3}}\frac{\sup|\Omega'|}{\epsilon_*}}.
 \label{eq:kad}
\end{equation}
For the representative benchmark we find $\sup|\Omega'|\simeq 1.610$, which gives $k_{\rm ad}(1)\simeq 0.787$, $k_{\rm ad}(0.1)\simeq 2.49$, and $k_{\rm ad}(0.01)\simeq 7.87$. The resulting bound is shown in \cref{fig:softbound}. The important consequence is that any genuinely nonadiabatic effect in the soft channel is confined to an IR band controlled directly by the symmetry-breaking rate.

\subsection{Quartic stabilization of the soft quintet}

A final structural question remains: is the soft homogeneous quintet actually unstable before symmetry breaking, or merely quadratically flat? The homogeneous matrix potential provides the answer. For a pure traceless-symmetric fluctuation $T$ in the unbroken phase ($q=0$, $\Omega=0$), one has
\begin{equation}
V_T=\frac{g^2}{4}\left[(\Tr T^2)^2-\Tr(T^4)\right].
\label{eq:VTraw}
\end{equation}
Because $T$ is a real symmetric traceless $3\times 3$ matrix, it may be diagonalized with eigenvalues $(\lambda_1,\lambda_2,\lambda_3)$ satisfying $\lambda_1+\lambda_2+\lambda_3=0$. One then finds the identity
\begin{equation}
\Tr(T^4)=\frac{1}{2}(\Tr T^2)^2,
\label{eq:tridentity}
\end{equation}
which reduces the potential to
\begin{equation}
V_T=\frac{g^2}{8}(\Tr T^2)^2\ge 0.
\label{eq:VTquartic}
\end{equation}
Thus the quintet is gapless at quadratic order but quartically stabilized. It is not tachyonic. The soft-channel problem is therefore an IR/vacuum-definition issue, not an instability artifact.

This distinction is important for the physical interpretation. A tachyonic mode would imply that the isotropic background is itself unstable already at the homogeneous level. The quartically stabilized quintet instead shows that the unbroken-phase background possesses a marginal direction whose dynamics are more delicate than those of a standard gapped Gaussian mode. This is precisely why the finite-$k$ problem demands further work.

\section{Phenomenological implications and finite-$k$ consistency estimates}\label{sec:pheno}

\begin{figure}[t]
  \centering
  \includegraphics[width=\columnwidth]{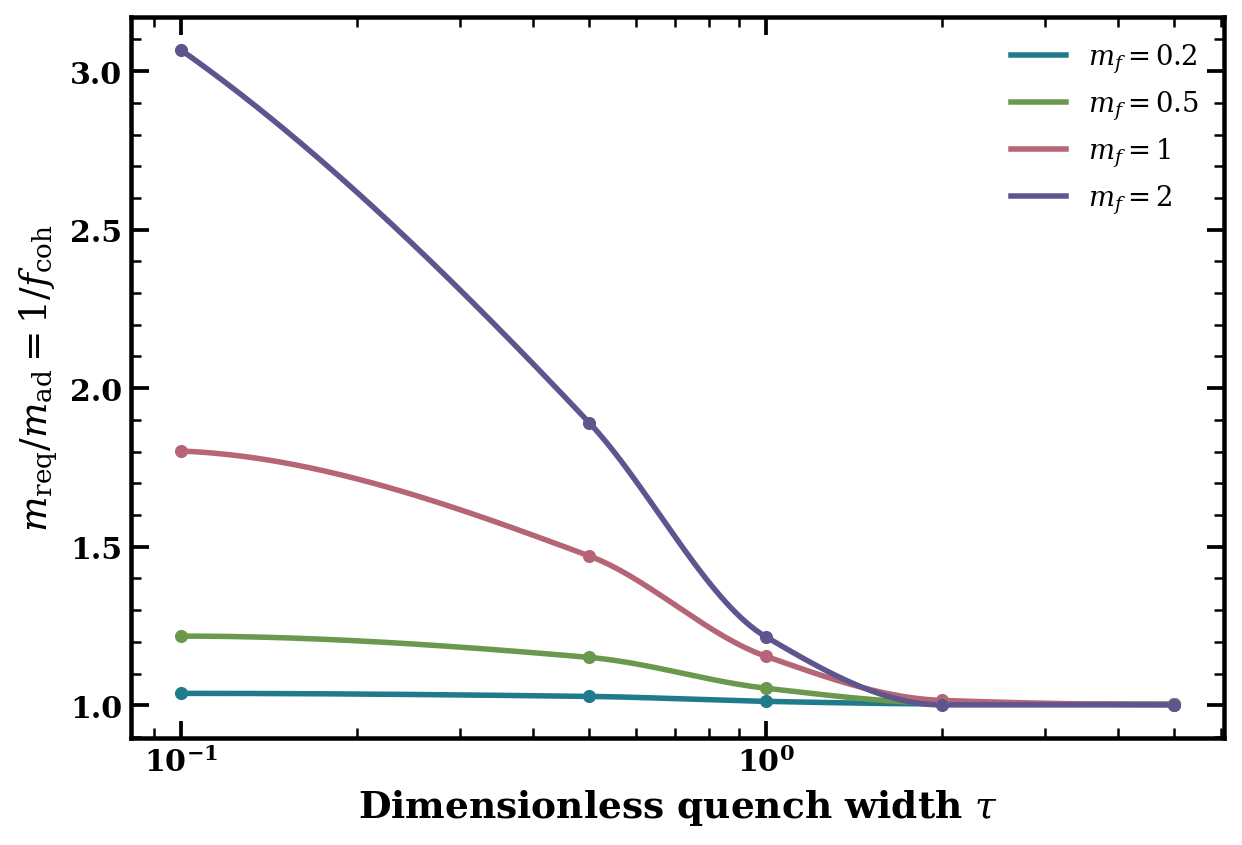}
  \caption{Representative mass-shift factor $m_{\rm req}/m_{\rm ad}=1/f_{\rm coh}$ as a function of the dimensionless quench width $\tau$ for several values of the asymptotic mass parameter $m_f$. The coherent-sector correction is small in the adiabatic regime and grows to order unity as the transition becomes faster.}
  \label{fig:massshiftcurves}
\end{figure}

\begin{figure}[t]
  \centering
  \includegraphics[width=\columnwidth]{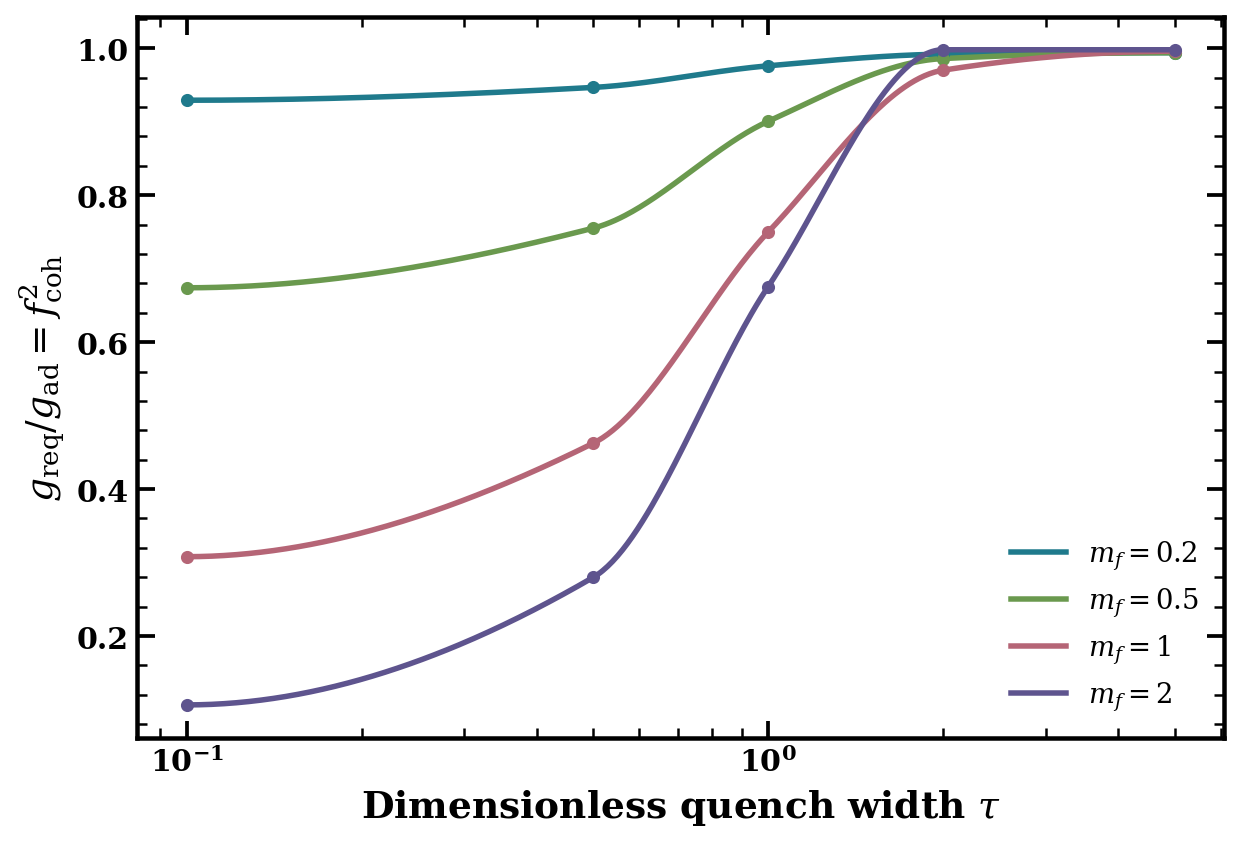}
  \caption{Representative coupling-shift factor $g_{\rm req}/g_{\rm ad}=f_{\rm coh}^2$ extracted from the same scan. The coherent-sector effect compresses the coupling normalization whenever the post-breaking transition is non-adiabatic.}
  \label{fig:gshiftcurves}
\end{figure}

The principal phenomenological output of the coherent-sector analysis is a controlled deformation of the post-transition matching problem. Equation~\eqref{eq:corrected_line} shows that any application of the inherited-condensate mechanism which relies on the coherent abundance must be interpreted through the survival factor $f_{\rm coh}$. In practice this means that the mass required at fixed $(g,c)$ and the coupling required at fixed $(m,c)$ are shifted by the simple factors
\begin{equation}
\frac{m_{\rm req}}{m_{\rm ad}}=\frac{1}{f_{\rm coh}},
\qquad
\frac{g_{\rm req}}{g_{\rm ad}}=f_{\rm coh}^2.
\end{equation}
The representative scans in \cref{fig:heatmap,fig:cohshift,fig:fcohscan,fig:massshiftcurves,fig:gshiftcurves} show that these corrections are mild once the transition is comfortably adiabatic, but can become order unity for sufficiently rapid smooth quenches.

The role of these figures is to isolate the coherent-sector response before a model-specific symmetry-breaking completion is imposed. The axes are the dimensionless parameters of the canonical quench problem: $m_f$ is the asymptotic mass scale measured in units of the local oscillation scale used in the numerical integration, and $\tau$ is the corresponding dimensionless transition width. A complete mapping back to the microscopic inflationary parameters requires a specific symmetry-breaking completion. The discussion of \cref{sec:numerics} shows how such a mapping may be organized in terms of the scalar relaxation rate $a_\mathrm{sb}m_\Phi$ in simple Higgs realizations, but the tanh scan should therefore be read as a model-independent response map rather than as the output of a unique ultraviolet completion.

This qualification is important because the same post-breaking abundance relation enters other phenomenological applications. Late-formation and small-scale-structure constraints depend on the epoch at which the inherited vector behaves as ordinary cold matter \cite{sarkar_das_sethi_2015,corasaniti_etal_2017,das_nadler_2021}. Since the coherent component is renormalized by the quench, the abundance contour entering those constraints is shifted before one addresses any inhomogeneous transfer. The effect should therefore be regarded as a dynamical correction to the coherent matching problem rather than as a replacement for the broader theoretical uncertainty associated with reheating or with the inflationary parameter $c$.

{A related issue concerns the fate of the energy that exits the coherent branch. The homogeneous analysis determines the survival factor $f_{\rm coh}$, while the finite-$k$ branching fractions into gauge, Higgs, and mixed channels require a dynamical calculation beyond the present matching problem. It is therefore useful to introduce an explicit relativistic branching parameter $0\le {\cal B}_{\rm rel}\le1$ and define
\begin{equation}
r_\ast^{\rm rel}\equiv {\cal B}_{\rm rel}(1-f_{\rm coh})R_{Q,\ast}^{\rm ad},
\label{eq:rrelbook}
\end{equation}
where $R_{Q,\ast}^{\rm ad}$ denotes the dark-to-visible energy ratio carried by the adiabatic coherent branch at the epoch when the relativistic channel is established. The envelope ${\cal B}_{\rm rel}=1$ corresponds to the conservative case in which all energy missing from the coherent branch remains in decoupled relativistic dark-sector quanta; smaller values describe conversion into massive vectors, Higgs excitations, turbulence, or other nonrelativistic dark-sector components. In this notation the dark-radiation contribution scales as
\begin{equation}
\Delta N_{\rm eff}\propto r_\ast^{\rm rel}.
\end{equation}
}

\subsection{A conservative finite-$k$ consistency estimate}

The regulated adiabatic bound of \cref{sec:soft} can be turned into a conservative consistency estimate for the unresolved inhomogeneous sector. For the minimal regulated soft channel one has \eqref{eq:kad}, so modes with $k\gtrsim k_{\rm ad}$ are guaranteed adiabatic while potentially nonadiabatic production is confined to the infrared band $k\lesssim k_{\rm ad}$. If one adopts the deliberately conservative assumption that the five quintet degrees of freedom are populated with order-unity occupation throughout that entire band at the epoch $\eta_*$ when the bound is saturated, the associated relativistic energy density is bounded by
\begin{equation}
 \rho_{T,*}^{\rm max} \sim \frac{5}{2\pi^2 a_*^4}\int_0^{k_{\rm ad}} \dd k\,k^3 = \frac{5}{8\pi^2}\frac{k_{\rm ad}^4}{a_*^4}.
\label{eq:rhoTmax}
\end{equation}
Comparing with the coherent energy density at the same epoch, $\rho_{{\rm coh},*}\simeq (3/2)m_*^2Q_*^2=(3/2)\Omega_*^2X_*^2/a_*^4$, gives the bound
\begin{equation}
 r_{T,*}^{\rm max}\equiv \frac{\rho_{T,*}^{\rm max}}{\rho_{{\rm coh},*}} \lesssim \frac{5}{12\pi^2}\frac{k_{\rm ad}^4}{\Omega_*^2X_*^2}.
\label{eq:rTmax}
\end{equation}
This estimate is deliberately conservative, but it already sharpens the regime of validity of the homogeneous treatment. The unresolved finite-$k$ sector becomes quantitatively important only if the infrared production band is broad enough and the coherent condensate is simultaneously small enough that \eqref{eq:rTmax} approaches unity. Conversely, whenever $r_{T,*}^{\rm max}\ll1$ the homogeneous survival factor is self-consistent as the leading correction to the post-transition coherent abundance.

{A related bookkeeping estimate may be written for late dark radiation. For a decoupled relativistic component with energy ratio $r_*^{\rm rel}$ at the epoch $T_*$, the contribution to the effective neutrino number is
\begin{equation}
\Delta N_{\rm eff} \simeq \frac{43}{7}\left(\frac{g_{*s}(T_\nu)}{g_{*s}(T_*)}\right)^{4/3}r_*^{\rm rel} .
\label{eq:dNeffestimate}
\end{equation}
Near the electroweak scale, $g_{*s}(T_*)\simeq 106.75$ gives $\Delta N_{\rm eff}\simeq 0.027\,(r_*^{\rm rel}/10^{-2})$. Equation~\eqref{eq:dNeffestimate} assumes a component that is decoupled from the visible sector, remains relativistic until the epoch at which $\Delta N_{\rm eff}$ is evaluated, and receives no later entropy injection, decay, or conversion into nonrelativistic matter. Additional contributions from direct finite-$k$ production, Higgs excitations, turbulence, or a pre-existing dark bath enter $r_*^{\rm rel}$ additively. The conservative envelope ${\cal B}_{\rm rel}=1$ in \eqref{eq:rrelbook} gives the maximum dark-radiation contribution associated with coherent depletion; for ${\cal B}_{\rm rel}<1$, the result scales linearly.
}

This is precisely where the soft-quintet analysis becomes relevant. The gap hierarchy, the exact cubic selection structure, the zero-mode obstruction, the regulated adiabatic bound, and the quartic stabilization together suggest that the soft quintet can act as a bottleneck rather than as an immediately democratic sink. At the same time, the gapless nature of this channel in the unbroken phase means that finite-$k$ excitations could in principle erode the homogeneous condensate before the mass turn-on is complete. In this sense the survival factor obtained from the coherent quench problem should be interpreted conservatively: it characterizes the renormalization induced by the non-adiabatic symmetry-breaking transition of the coherent branch itself, while any efficient pre- or mid-transition transfer into inhomogeneous modes would only reduce the coherent remnant further. The coherent-sector result therefore provides a natural upper envelope for the surviving homogeneous abundance prior to a full finite-$k$ helicity-resolved calculation.

Accordingly, the present work should be viewed as a completed coherent-sector analysis together with a structural map of the unresolved fluctuation problem. That already suffices to reinterpret the adiabatic matching branch, to quantify when the post-transition correction is parametrically negligible, and to identify the infrared window in which any complete gauge-Higgs transfer calculation must concentrate.

\section{Conclusions}\label{sec:conclusions}

We have tackled the post-inflationary symmetry-breaking puzzle for inherited axion–SU(2) condensates with a clear target: push the coherent transition as far as possible before confronting the full finite‑$k$ gauge–Higgs problem. The framework we built is both compact enough to be practical and precise enough to force a rethink of how the abundance‑matching problem should be posed.

At the level of the spatially homogeneous condensate, introducing the canonical conformal-time variable $X = aQ$ reformulates the transition dynamics as those of an oscillator with a time-dependent quartic-plus-quadratic potential. The corresponding action variable furnishes a single invariant quantity that remains valid across the crossover from quartic to quadratic dominance and thereby determines the adiabatic matching coefficient,
\begin{equation}
C_{\rm ad} = \frac{2\Gamma(1/4)}{3\sqrt{\pi}\,\Gamma(3/4)}.
\end{equation}
The relations governing the quench work and the excess energy subsequently imply that the conventional abundance formula is most appropriately interpreted as the zero–excess-energy branch of the post-inflationary quench problem. In the general case, the coherent abundance is renormalized by the survival factor $f_{\rm coh} = J_{\rm late}/J_{\rm early}$, and continuous yet sufficiently rapid transitions can suppress the coherent component at order unity.

At the level of homogeneous fluctuations, the matrix formulation leads to a diagonal-$SO(3)$ decomposition into $1\oplus 3\oplus 5$ channels and isolates a traceless-symmetric quintet whose quadratic gap is supplied only by symmetry breaking. The corresponding cubic interaction network is selective rather than democratic, and the soft channel exhibits a $k=0$ vacuum obstruction, a regulated ultraviolet adiabaticity bound, and quartic stabilization. These structural results do not close the finite-$k$ problem, but they identify the infrared organization that any complete treatment must reproduce.

The bottom line is sharp. The inherited-condensate picture does not stop at adiabatic matching; it comes with a concrete, post-transition renormalization whose magnitude is dictated by the tug-of-war between the oscillation timescale and the symmetry-breaking timescale. That single insight is enough to redraw abundance contours, to pinpoint when the standard matching rule can be trusted, and to spotlight the slice of parameter space where additional inhomogeneous dynamics are poised to have the greatest impact.

Several natural extensions of this work suggest themselves. A helicity-resolved, finite‑$k$ gauge–Higgs computation would elucidate how the coherent depletion channel bifurcates into propagating excitations, while preserving the diagonal‑$SO(3)$ structure identified in the present analysis. A fully nonlinear treatment of the transition regime—for instance, employing Hartree‑resummed dynamics or lattice simulations—would allow a quantitative characterization of energy redistribution, dark radiation production, and the onset of post‑symmetry‑breaking turbulence in explicit Higgs-sector completions. Since the coherent sector renormalizes independently of the inflationary initial conditions, these extensions can be implemented on top of the current framework without modifying its matching procedure. {The thermal benchmark used here is an order-parameter realization of a mass quench in the kinetically resolved coherent regime, rather than a fully thermalized $SU(2)$--Higgs plasma. The dark-radiation formula is a conversion from a specified relativistic branching fraction to $\Delta N_{\rm eff}$; the branching fraction itself is fixed by the finite-$k$ gauge--Higgs dynamics identified above as the next stage of the program.}

\appendix

\section{Derivation of the homogeneous matrix potential}\label{app:matrix}

For a homogeneous gauge matrix $X_i{}^a(\eta)$, the Yang--Mills field strength in temporal gauge contains only the electric component $X'_i{}^a$ and the homogeneous non-Abelian magnetic term proportional to $g\epsilon^{abc}X_i{}^bX_j{}^c$. After contraction of spatial and gauge indices one obtains the homogeneous Lagrangian quoted in \eqref{eq:matrixL}. The quartic invariant may be written in terms of the positive matrix $M=XX^T$ as
\begin{equation}
V_4=\frac{g^2}{4}\left[(\Tr M)^2-\Tr(M^2)\right].
\end{equation}
For the isotropic background $X=q\,\mathbb{1}$, one has $M=q^2\mathbb{1}$ and therefore
\begin{equation}
V_4=\frac{g^2}{4}\left[(3q^2)^2-3q^4\right]=\frac{3}{2}g^2q^4,
\end{equation}
recovering the quartic term of the canonical background Lagrangian.

\section{Quadratic and cubic expansion around the isotropic background}\label{app:cubic}

Substituting $X=q\,\mathbb{1}+\chi\mathbb{1}+A+T$ into \eqref{eq:matrixL} and expanding to second order gives \eqref{eq:L2channels}. The cubic potential is obtained most efficiently by symbolic expansion of the invariants and reads
\begin{equation}
V_3 = g^2 q\Big[6\chi^3+2\chi\Tr(AA^T)-\Tr(T^3)-\Tr(TAA^T)\Big].
\end{equation}
The quadratic and cubic identities quoted in the main text follow from direct expansion of the matrix invariants and are displayed here in analytic form.

\section{Quartic identities for traceless symmetric $3\times3$ matrices}\label{app:quartic}

Let $T$ be a real symmetric traceless $3\times3$ matrix. Since $T$ is orthogonally diagonalizable, one may write $T={\rm diag}(\lambda_1,\lambda_2,\lambda_3)$ with $\lambda_1+\lambda_2+\lambda_3=0$. Then
\begin{equation}
(\Tr T^2)^2-\Tr(T^4)=2\sum_{i<j}\lambda_i^2\lambda_j^2.
\end{equation}
Using $\lambda_3=-(\lambda_1+\lambda_2)$, one finds after a short algebraic exercise that
\begin{equation}
\Tr(T^4)=\frac{1}{2}(\Tr T^2)^2.
\end{equation}
Substitution into \eqref{eq:VTraw} yields the positive quartic stabilization law \eqref{eq:VTquartic}.

\section{Illustrative mapping between the effective width and a Higgs transition}\label{app:higgs}

Although the main text treats the post-breaking mass profile as an effective interpolation, the width parameter admits a simple physical interpretation in a broad class of symmetry-breaking models. Consider a scalar order parameter $\Phi$ with late-time vacuum expectation value $v_\Phi$ and quartic potential
\begin{equation}
V(\Phi)=\frac{\lambda_\Phi}{4}\left(|\Phi|^2-v_\Phi^2\right)^2.
\end{equation}
If the gauge bosons acquire their mass through $m(\eta)=g\,v(\eta)$ with $v(\eta)=\sqrt{2\langle|\Phi|^2\rangle}$, then near the minimum the scalar relaxation scale is set by
\begin{equation}
m_\Phi^2\simeq 2\lambda_\Phi v_\Phi^2,
\end{equation}
so the conformal transition width scales as
\begin{equation}
\tau_\Phi\sim (a_\mathrm{sb}m_\Phi)^{-1}
\end{equation}
up to order-unity corrections from Hubble damping and from the details of the initial displacement. The dimensionless parameter $\tau$ used in the main text therefore probes the ratio of this scalar relaxation time to the characteristic oscillation time of the inherited condensate. Rapid quenches correspond to scalar relaxation on timescales comparable to or shorter than the condensate oscillation period, while slow quenches correspond to a scalar evolution that is adiabatic relative to that period.

This mapping is sufficient for the present purpose: it explains how a concrete symmetry-breaking sector feeds into the canonical quench parameters without forcing the analysis into one special Higgs completion. A dedicated study of the coupled scalar-gauge dynamics would be required to turn this illustrative mapping into a fully predictive scan.

\section{Numerical implementation and convergence diagnostics}\label{app:numerics}

The numerical results in \cref{sec:numerics} are obtained from explicit integration of the local quench equation \eqref{eq:local_quench_eq} and the FRW canonical equation \eqref{eq:Xeq}. In both cases we employ adaptive Runge--Kutta integration with relative tolerances between $10^{-8}$ and $10^{-9}$ and absolute tolerances between $10^{-10}$ and $10^{-11}$. Late-time amplitudes are extracted from the average of the last ten turning points once the solution has entered the asymptotic oscillatory regime. For the FRW action plots the turning-point action is evaluated from the unified invariant \eqref{eq:J_unified} at each turning point.

For the local quench benchmark, the quantities displayed in \cref{fig:localtraj,fig:localconv} are stable at the sub-$10^{-3}$ level under simultaneous tightening of the numerical tolerances and under moderate changes in the extraction window for the late-time envelope. For the FRW benchmark, the relative RMS agreement between the numerically differentiated Hamiltonian and the work identity \eqref{eq:energy_identity} is at the level quoted in the main text, and the difference between the direct excess-energy extraction and the late-quadratic closure formula remains below one percent in the benchmark run. These diagnostics indicate that the figures are controlled by the physical transition rather than by numerical noise.

The phenomenological plots should be interpreted with the same care as the underlying scan. They are intended to illustrate the coherent-sector response of the post-breaking condensate as the transition width and asymptotic mass are varied within the canonical quench problem. They are not substitutes for a complete microphysical parameter scan of a specific symmetry-breaking completion.

\bibliographystyle{apsrev4-2}
\bibliography{refs}

\end{document}